# Usage habits of business information system in Hungary

Peter Sasvari, Institute of Business Sciences, University of Miskolc, Hungary

*Abstract*— The IT functions of the companies can be executed in different ways in-house solution, outsourcing, in sourcing, formation a spin-off company. Predominantly this function is provided within the company in Hungary. The larger a company is; it is more likely that a separate IT manager will be entrusted for the supervision of IT functions. Only a very small number of small-sized enterprises said that they paid special attention to formulating an IT strategy, while it was not considered important by microenterprises at all.

## I. INTRODUCTION

The role of information has become more and more substantial in the economy recently, and information is regarded as an important resource since it is more difficult for companies to improve their market positions in the long term without having the appropriate amount of available information. Globalization in the business world has brought about the possibility of getting a greater amount of information in much less time which means that companies are forced to spend more time and energy on handling the increased information load. [1] Business information systems are designed to provide effective help in this process as they are becoming increasingly popular among companies due to the robust technological development. This paper deals with the usage of business information systems among the Hungarian enterprises and analyzes the following three key questions: how the usage of business information systems influences a company's economic performance, how much is the expenditure for an individual company to develop its information technology infrastructure and finally, to what extent information technology is considered important as a functional area within the organization of a company. [2] The aim of the research presented in this paper was to explore the current situation of Hungarian enterprises in terms of using business information systems, gaining a more thorough insight into the background of the decisions made on introducing such information systems together with the possible problems related to their introduction and further usage.

## II. THE BACKGROUND LITERATURE OF BUSINESS INFORMATION SYSTEMS

There are several definitions offered on business information systems in the literature. According to Burt and Taylor's approach, "business information systems can be regarded as an information source in any combination thereof, or any access to and any recovery of their use or manipulation. Any business information system is designed to link the user to an appropriate source of information that the user actually needs, with the expectation that the user will be able to access the information satisfying their needs" [3]. Davis and Olson define business information systems as "an integrated user-machine system for providing information to support the operations, management, analysis, and decision-making functions in an organization. The system utilizes computer hardware and software, manual procedures, models for analysis, planning, control, and decision-making by using a database" [7]. "Information systems are a part of any organization that provides generates, stores, separates divides and uses information. They are made up of human, technical, financial and economic components and resources. In fact, they can be regarded as inherently human systems (organizations, manual systems) that may include a computer system, and automatizes certain well-defined parts and selected items of the system. Its aim is to support both the management functions and the daily operation of an organization." [8] In a broader sense, a business information system is the collection of individuals, activities and equipment employed to collect, process and store information related to the company's environment, its internal activities, together with all transactions between the company and its environment. Beyond giving direct support to operations, its basic task is to provide decision-makers with the necessary information during the whole decision-making process. The system's main components are the following [10]:

- Individuals carrying out corporate activities: the actual users of technical apparatus. Decision-makers also belong to this group, as leaders who receive information on the factors affecting business operations, and use business information systems to make decisions in relation to planning, implementation and monitoring business activities.
- Information (also known as processed data on external and internal facts) which – due to its systematized form – can be used directly in the decision-making process.

[1] The described work was carried out as part of the TÁMOP-4.2.1.B-10/2/KONV-2010-0001 project in the framework of the New Hungarian Development Plan. The realization of this project is supported by the European Union, co-financed by the European Social Fund.

- Technical apparatus, nowadays usually a computer system that supports and connects the subsystems applied to achieve corporate objectives.

The computer system standardizes a significant part of the information and communication system, thus making it easier to produce and use information. According to one definition proposed [5] "information systems are systems that use information technology to collect information, transmit, store, retrieve, process, display and transform information in a business organization by using information technology." Raffai's understanding of information systems is as follows: "it uses data and information as a basic resource for different processing activities in order to provide useful information for performing useful organizational tasks. It's main purpose is the production of information, that is dedicated to creating messages that are new to the user, uncertainties persist, and their duties, to assist in fulfilling the decisions" [19]. The classification of business information systems is a difficult task because, due to the continuous development, it is hard to find a classification system that can present unanimously defined information system types. It occurs quite often that different abbreviations are used to refer to the same system or certain system types appear to be merged together. As a consequence, the classification of business information systems can be done in several ways, the lists of several groups of business information systems presented below just to show a few alternatives for classification [2]. Dobay [8] made a distinction between the following types:

- Office Automation Systems (OAS): used for efficient handling of personal and organizational data (text, image, number, and voice), making calculations and document management.
- Communication Systems: supporting the information flow between groups of people in a wide variety of forms.
- Transaction-Processing Systems (TPS): used for receiving the initiated signals of transactions, generating and giving feedback on the transaction event.
- Management Information Systems (MIS): used for transforming TPS-related data into information for controlling, management and analysis purposes.
- Executive Information Systems (EIS): intended to give well-structured, aggregated information for decision-making purposes.
- Decision Support Systems (DSS): applied to support decision-making processes with information, modeling tools and analytical methods.
- Facility Management Systems (facility management, production management): used for directly supporting the value production process.
- Group Work Systems: intended to give group access to data files, to facilitate structured workflows and the implementation of work schedules.

Another possible approach to defining categories is based on Raffai's work [19]:

- Implementation support systems: this group includes transaction processing systems (TPS), process control systems (PCS), online transaction processing systems (OLTP), office automation systems (OAS), group work support systems (GS), workflow management (WF), and customer relation management systems (CRM).
- Executive work support systems: this category can include strategic information systems (SIS), executive information systems (EIS), online analytical processing systems (OLAP), decision support systems (DSS), group decision support systems (GDSS), and management information systems (MIS).
- Other support systems: business support systems, (BIS), Expert Systems (ES), integrated information processing systems (IIS), and Inter-Organizational Information Systems (IOS) can be found in this category.

When a business organization makes a decision about introducing any business information system, their decision can be explained by a variety of factors. The most common factors are as follows [15]:

- "Technical considerations: companies applying fragmented, outdated business information systems with the lack of transparency.
- Strategic considerations: ERP systems may play a role in maintaining and enhancing competitiveness, they may establish the technical background to apply e-commerce solutions.
- Business considerations: among others, cost reduction and profit increase objectives, job cuts, stock reduction, reducing IT costs, improving productivity and more rapid turnaround of orders may belong to this group of factors."

Ideally, before a company decides to introduce a business information system, they consider a large number of factors. The most important step during this process is to select the most relevant aspects, then, after weighing them carefully, the management of a company can choose the best offer available.

### III. THE AIM AND THE CONCEPT OF THE RESEARCH

The review of the relevant literature on the subject made it possible to identify the most important points of the research. Based on these, the main objectives as well as the concept of the research were formulated. The research objectives are the following:

- To present the background of the decisions related to the introduction of business information systems, along with the problems encountered in the phase of their introduction,
- To analyze the usage patterns of business information systems,
- To reveal the connection between using business information systems and the operational effectiveness or profitability of companies.

Based on the aims presented above, the following research concept was determined:
- First, the major issues related to the introduction of business information systems were analyzed. It was surveyed whether the companies taking part in the study used any sort of business information systems, and if not, what causes or conditions prevented them from introducing them. In the case of companies applying business information systems, the causes of introducing such systems, the information sources for selecting the appropriate systems, and the criteria for selecting them were also investigated. It was also examined whether companies had made calculations on the costs of introducing business information systems before making decisions on them, and if so, what aspects had been taken into account during the calculation. The problems occurring in the phase of implementation were identified.
- After that, the usage patterns of business information systems at companies were examined. The main points of the relevant analysis were the given company's information technology infrastructure, its Internet usage habits, and its appearance on the Internet. Here, the types of the applied business information systems and their areas of use were also presented, then the forms of information technology functions, human resource issues and the main points of IT strategy were covered.
- In the closing part of the analysis, the impacts of using business information systems on the operational effectiveness of companies were examined. It was investigated whether the introduction of such systems had an influence on the performance and revenues of the company as well as the size of the targeted market and the changes on the demand side. Another point of the investigation was whether ensuring more efficient information flow and information management contributed to the reduction of the company's other costs. These factors, summarized in Figure 1, of course, cannot be quantified so easily; however, taking them into consideration can lead to making more firmly grounded decisions on the introduction of various business information systems.

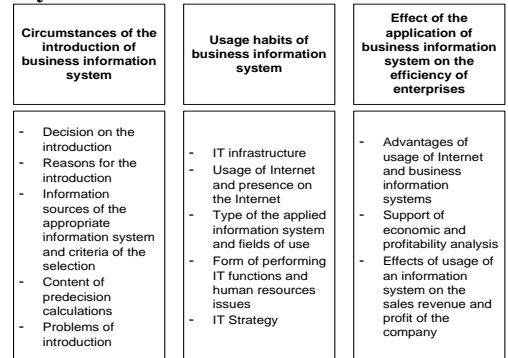

**Fig 1 - Conceptual Model of the Primary Research**

### IV. THE RESEARCH METHOD

The empirical survey was carried out using a written questionnaire. In the phase of compiling the individual questions of the survey, the main results of the previously conducted empirical surveys on the subject were also taken into account. The questionnaire was divided into five major parts. The first part included some basic questions about the companies' background (such as their location, fields of operation, number of employees etc.), then questions related to the responding company's information technology infrastructure followed. In the third part of the questionnaire, the emphasis was put on questions enquiring about the Internet-using habits of the companies; the fourth group of questions was aimed at enquiring about the usage patterns of business information systems, making it the most detailed part of the questionnaire. The closing part contained questions about the IT-skilled human resources employed by the responding companies. The questionnaire was sent out to several hundreds of companies, The Hungarian survey was conducted both in a paper-based format and online with the assistance of the software application called Evasys. For evaluating data and presenting the results of the survey, the statistical software packages Excel 2007 and SPSS 19.0 were applied. The 21% of the Hungarian responder companies are micro-sized, 29% are small-sized, 29% are medium-sized enterprise and 21% are corporations.

### V. THE USAGE PATTERNS OF BUSINESS INFORMATION SYSTEMS

The IT functions of the companies can be executed in different ways:
- in-house solution,
- outsourcing,
- insourcing,
- Formation a spin-off company.

For the question that in what form companies solve the provision of IT functions, the following results were obtained (see Figure 2). Predominantly this function is provided within the company; however this solution is applied among small-sized enterprises at a slightly lower rate. In contrast, outsourcing mainly occurs among small-sized enterprises, more than one-third of them plan to outsource the provision of their IT functions. This sounds logical, as at small-sized enterprises an appropriate person is not always available for the provision of their IT functions. The spin-off, which means a formation of a non-independent, non-separate company from the ownership point of view, is a slightly used solution and can be observed only among small and medium-sized enterprises. None of the respondents took into account the possibility of insourcing. Based on the responses given in the „Other" category, it can be stated that certain companies apply more from the available possibilities at the same time and beside the in-house function provision, the central IT department of the foreign parent company also helps in managing IT functions.

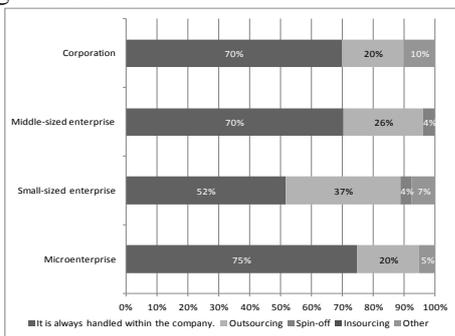

**Fig 2 Forms of the Provision of IT Functions by Company Size**

The next question was about determining who was responsible for the operation of the IT functions within the company (see Figure 3). It can be stated that the vast majority of the corporations place great importance on having a separate IT manager being in charge of this field. Sometimes the assignment of a third company happens or a lower-level employee is appointed to fulfill tasks related to IT functions, but it could also be seen that the company's top manager at any corporation was not responsible for the operation of this function. There were some surprising responses coming from microenterprises in the „Other" category, namely according to some companies a „knowledgeable family member" was responsible for their IT functions, or nobody was appointed to deal with this field.

**TABLE 1. PHI AND CRAMER'S V (THE RESPONSIBLE PERSON FOR THE OPERATION)**

| The responsible person for the operation | Phi | | Cramer's V | |
| --- | --- | --- | --- | --- |
| | Value | Approximate Significance | Value | Approximate Significance |
| **Chief Information Officer or Chief Executive Officer or An appointed employee working at a departmental level or An external service provider** | 0.718 | 0.000 | 0.415 | 0.000 |

The studies showed that there was a strong relationship between the company size and the person responsible for IT functions. The larger a company is; it is more likely that a separate IT manager will be entrusted for the supervision of IT functions. (see Table 1)

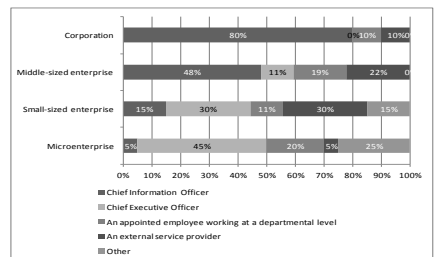

**Fig 3 the Responsible Person for the Operation of the IT Function by Company Size**

The next question concerning IT-related human resource was: „Did your company employ any IT professionals in 2010?". The results are as follows. By definition, an IT professional employee is a person who has qualification for the planning, development, deployment, operation, support, management and evaluation of IT systems. Figure 4 clearly shows that corporations can afford the employment of IT professionals (90%) and as the company size becomes smaller, the employment of separate IT professionals is less typical for the supervision of IT tasks.

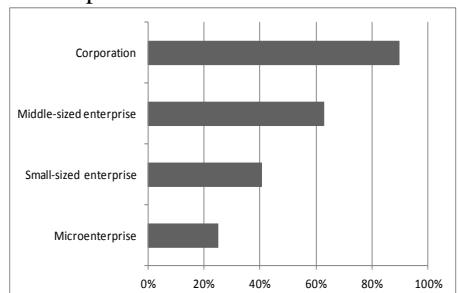

**Fig 4 - Employment of IT Experts by Company Size**

As it is indicated by the received responses (see Figure 5), mainly corporations tended to formulate independent IT strategies but this issue was also dealt with as a part of corporate strategy. However, about a third of the corporations do not formulate any kind of IT strategy and they do not even have such plans for the future. A much lower rate, only 15 % of the surveyed medium-sized companies indicated that they were in the making of IT strategies, but, compared to corporations, a greater proportion of them said that the issues related to IT strategy were covered by their overall corporate strategy. Only a very small number of small-sized enterprises said that they paid special attention to formulating an IT strategy, while it was not considered important by microenterprises at all.

**Table 2 Phi and Cramer's V (The Practice Of Formulating IT Strategy Based)**

| The practice of formulating IT strategy based | Phi | | Cramer's V | |
| --- | --- | --- | --- | --- |
| | Value | Approximate Significance | Value | Approximate Significance |

| 'Yes, we formulate a separate information strategy' or 'Yes, we take it as a part of our broader corporation strategy' or 'No, we do not but it is under planning' or 'No, we do not and we do not plan it, either' | 0.574 | 0.000 | 0.331 | 0.000 |

A significant but weaker-than-average correlation was shown between company sizes and formulating IT strategies. This clear correlation may be due to the fact that the structure and the resources of a corporation makes it more reasonable for them to formulate an independent IT strategy than it could be seen in the case of microenterprises. (see Table 2)

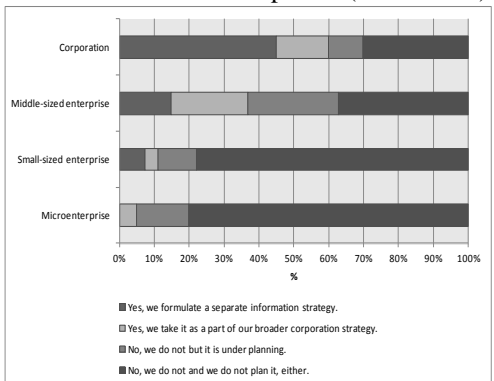

**Fig 5 – The Practice of Formulating IT Strategy Based On Company Size**

The companies dealing with the preparation of IT strategy on some level also made a declaration on its content. The results of this are shown in Figure 6. 100% of companies preparing an independent IT strategy also deal with the technical aspects of their IT strategy. A part of this, they pay attention to the concept of servers, workstations and network development. 92 % of them also mentioned the beneficial business aspects of IT, for instance such applications that provide competitive advantages or support cost-benefit analyses. Companies in a slightly lower ratio (67%) deal with the management of IT functions. Within this question, they could decide on the place of internal IT organization, accounting of the related costs and on the application of external service providers. Regarding the content of the applied IT strategies, there are large differences concerned by size categories. The fields mentioned above appear in a significantly smaller ratio or do not appear in the SME's IT strategy at all. Besides the above contents, the respondents were allowed to list "Other" affected areas as well. The following answers were given to this open question: "application and device consolidation, reducing the numbers of the systems, increase of integration, increase physical security (backup) of the operation and a safe storage of information".

TABLE 3 – PHI AND CRAMER'S V (IT STRATEGY)

| IT strategy | Phi | | Cramer's V | |
|---|---|---|---|---|
| | Value | Approximate Significance | Value | Approximate Significance |
| **Technical issues (network development along with servers and workstations)** | 0.539 | 0.056 | 0.539 | 0.056 |
| **Business aspects of information technology (applications securing competitive edge, cost-benefit analysis)** | 0.648 | 0.012 | 0.648 | 0.012 |
| *Information system management (the place of IT within the organization, payment administration, contracting with external service providers)* | *0.465* | *0.132* | *0.465* | *0.132* |

Connecting to the above-listed answers, a close relationship could also be observed between the size of the company and the content of their information strategy. Regarding the technical issues and the management of IT functions, the relationship is weaker than average, while concerning the business implications of the information technology, the relationship showed medium strength. (see Table 3)

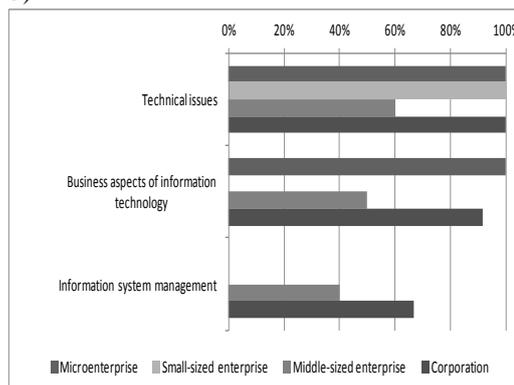

**Fig 6 - Contents of the IT Strategy**

## VI. CONCLUSION

Nowadays the issue of information technology in business is moving into the centre of attention, which is also indicated by the fact that more and more companies, not accidentally, recognize its importance. My aim was to present the circumstances of the decisions made about the introduction of business information systems and problems emerging during the introduction as well as to analyze the usage habits of companies applying these systems, and to explore the relation between the application of business information systems and the operational effectiveness of the business. Based on the scientific literature, I worked out a conceptual model appropriate for the aims of the research, serving as a base both for the questionnaire and the analysis. The primary focus of the analysis was to explore the differences and similarities of the usage habits of business information system by size categories. Thus, the micro-, small and medium-sized enterprises as well as corporations were also presented in the sample.

## REFERENCES

[1] M. Aranyossy, "Business Value of IT Investment: The Case of a Low Cost Airline's Website". In: 20th Bled eConference eMergence: Merging and Emerging Technologies, Processes, and Institutions, June 4 - 6, 2007, Bled, Slovenia.

[2] B. Bencsik, "Az üzleti információs rendszerek használati szokásainak elemzése a vállalkozások körében (Analysis of the usage practice of business information systems among the enterprises)", Szakdolgozat (MSC Thesis), Miskolc, 2011.

[3] E. Burt, and John A. Taylor, "Information and Communication Technologies: Reshaping Voluntary Organizations?", Nonprofit Management and Leadership, Volume 11, Issue 2, pages 131–143, Winter 2000, 2003.

[4] P. Csala, A. Csetényi, and B. Tarlós, "Informatika alapjai (Basis of informatics)", ComputerBooks, Budapest, 2003.

[5] L. Cser, and Z. Németh, "Gazdaságinformatikai alapok (Basis of economic informatics)", Aula Kiadó, Budapest, 2007.

[6] G. B. Davis, and M. H. Olson, "Management information systems: Conceptual foundations, structure, and development",. New York: McGraw-Hill, 1985.

[7] I. Deák, P. Bodnár, and G. Gyurkó, "A gazdasági informatika alapjai (Basis of economic informatics)", Perfekt Kiadó, Budapest, 2008.

[8] P. Dobay, "Vállalati információmenedzsment (Corporate information management)", Nemzeti Tankönyvkiadó, Budapest, 1997.

[9] G. Drótos, K. Gast, P. Móricz, and G. Vas, "Az információmenedzsment fejlettsége és a versenyképesség. Versenyben a világgal 2004-2006 gazdasági versenyképességünk vállalati nézőpontból c. kutatás. Versenyképesség kutatások műhelytanulmány-sorozat (State of development of information management and competitiveness. Research titled of 'Competition with the world, our economic competitiveness from corporate point of view between 2004 and 2006')". 28. sz. műhelytanulmány. Budapest, 2006.

[10] F. Erdős, "A kis- és közepes vállalkozások versenyképességének növelése integrált vállalatirányítási rendszerek által (Increase of the small and middle-sized enterprises' competitiveness by integrated business management systems)". Széchenyi István Egyetem, 2005.

[11] Gy. Fülöp and G. I. Pelczné, "The SME-Sector Development Strategy in Hungary", Global Management World Conference, Porto, Portugal, 2008.

[12] A. Gábor, "Üzleti informatika (Business informatics)", Aula Kiadó, Budapest, 2007.

[13] C. Harland, "Supply Chain Management, Purchasing and Supply Management, Logistics, Vertical Integration, Materials Management and Supply Chain Dynamics", Blackwell Encyclopedic Dictionary of Operations Management. UK: Blackwell, 1996.

[14] J. Hughes, "What is Supplier Relationship Management and Why Does it Matter?", DILForientering, 2010.

[15] L. Kacsukné Bruckner and T. Kiss, "Bevezetés az üzleti informatikába (Introduction into business informatics)". Akadémiai Kiadó, Budapest, 2007.

[16] P. Laudon, "Management Information Systems: Managing the Digital Firm", Prentice Hall/CourseSmart, 2009.

[17] A. Nemeslaki, "Vállalati internetstratégia (Corporate Internet Strategy)", Akadémiai Kiadó, Budapest, 2012.

[18] J. O'Brien, "Management Information Systems – Managing Information Technology in the Internetworked Enterprise", Boston: Irwin McGraw-Hill, 1999.

[19] M. Raffai, "Információrendszerek fejlesztése és menedzselése (Development and management of information systems)". Novadat Kiadó, 2003.

[20] P Sasvari, "A Conceptual Framework for Definition of the Correlation Between Company Size Categories and the Proliferation of Business Information Systems in Hungary", Theory, Methodology, Practice, Club of Economics in Miskolc, Volume 8: 2012, P 60-67, 2012.

[21] R. Shaw, "Computer Aided Marketing and Selling", Rbhp Trade Group, ISBN 978-0750617079, 1991.

[22] http://www.gazdasag.sk/szk-gazdasagi-fejlodes-mutatok,downloaded: 2012.10.22.

[23] Statisztikai tükör: A kis- és középvállalatok és a vállalkozás helyzete (http://www.ksh.hu/docs/hun/xftp/stattukor/kkv.pdf, downloaded: 2012.10.26.).

[24] http://www.portfolio.hu/gazdasag/doing_business_ot_helyet_csuszott_le_magyarorszag.174644.html, downloaded: 2012.10.30.

[25] http://users.iit.uni-miskolc.hu/ficsor/inftervseg/infrendsz1hand.pdf, downloaded: 2012.10.27.

[26] M. Szepesné Stiftinger, Rendszertervezés 1., Az információrendszer fogalma, feladata, fejlesztése (http://www.tankonyvtar.hu/hu/tartalom/tamop425/0027_RSZ1/ch01s03.html, downloaded: 2012.11.01.).

[27] http://www.gazdasag.sk, downloaded: 2012.10.27.

[28] Hungarian Central Statistical Office, http://www.ksh.hu, downloaded: 2012.10.27.

**AUTHOR'S PROFILE**

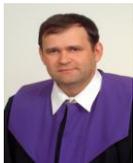

Dr. Peter Sasvari received his MSc in Mechanical Engineering, MSc in Economics and his PhD in Business and Organization Sciences at the University of Miskolc. Now he is an associate professor at the Institute of Business Science, Faculty of Economics, and University of Miskolc, Hungary. His current research interests include different aspects of Economics, ICT and the information society.